\documentclass[
    ,final            
, a4paper, tnotealph
  ]
  {aipproc}

\layoutstyle{6x9}

\newcommand {\hi} {{\rm H}\,{\small\rm I}}

\newcommand {\kmskpc} {\,{\rm km\,s}^{-1}\,{\rm kpc}^{-1}}
\newcommand {\de}{^{\circ}}
\newcommand {\mo}{\,{M}_\odot}

\newcommand{\K}{\,{\rm K}}
\newcommand {\moyr}{\,{M_\odot\,\rm yr}^{-1}}
\newcommand{\gsim}{\lower.7ex\hbox{$\;\stackrel{\textstyle>}{\sim}\;$}}
\newcommand{\lsim}{\lower.7ex\hbox{$\;\stackrel{\textstyle<}{\sim}\;$}}

\begin{document}

\title{Gas circulation and galaxy evolution}


\author{Filippo Fraternali}{
  address={Astronomy Department, University of Bologna, via Ranzani 1, 40127, Bologna (I)}
}


\begin{abstract}
Galaxies must form and evolve via the acquisition of gas from the intergalactic environment, however the way this gas accretion takes place is still poorly understood.
Star-forming galaxies are surrounded by multiphase halos that appear to be mostly produced by internal processes, e.g., galactic fountains.
However, a small fraction of the halo gas shows features that point to an external origin.
Estimates of the halo-gas accretion rate in the local Universe consistently give values much lower than what would be required to sustain star formation at the observed rate. 
Thus, most of the gas accretion must be ``hidden'' and not seen directly.
I discuss possible mechanisms that can cause the intergalactic gas to cool and join the star-forming galactic disks.
A possibility is that gas accretion is driven by the galactic-fountain process via turbulent mixing of the fountain gas with the coronal low-metallicity gas.

\end{abstract}

\maketitle


\section{Introduction}

Galaxy evolution is regulated by the ability of galaxies to accrete and retain cold gas, which is necessary for the process of star formation (SF) and fundamentally affects their morphologies.
Galaxies are broadly divided into ``early'' and ``late'' types or, in recent terminology, ``red sequence'' and ``blue cloud'' galaxies \cite{strateva+01}.
The former are actively star-forming while the latter have had their main SF episodes sometime in the past.
Blue and red galaxies are not found with the same ratio in all environments; red galaxies are more common in groups and clusters than in the field \cite{dressler80}.
On the other hand, galaxy scaling relations change only slightly with the environment suggesting that the mechanisms of galaxy formation and evolution are the same \cite{blanton+moustakas09}.
In the following, I investigate how galaxy evolution is influenced by the circulation of gas between relatively isolated disk galaxies and their environments.

The gas reservoir of galaxies must ultimately be the intergalactic medium (IGM), where most of the baryons are thought to reside \cite{fukugita+98}.
In a popular scheme for galaxy formation, gas falls onto dark matter potential wells and promptly shocks to the virial temperature \cite{white+rees78}.
Galaxies are thus thought to form from the cooling of such ``virialised'' hot atmospheres.
Recently, this picture has been challenged by new numerical simulations showing that a large fraction of gas can in fact reach the centre of the potential well in a ``cold'' form and stars can form out of this gas \cite{keres+05}.
This so-called ``cold mode'' of gas accretion, as opposed to the classical ``hot mode'', is supposed to take place in dark matter halos with relatively low masses $M_{\rm halo} < 10^{12} \mo$ \cite{dekel+birnboim06}.
Some simulations suggest that most SF occurs via this cold mode, which is dominant for all galaxies at redshifts $z \gsim 1$ \cite{keres+09}.

The main visible effect of gas accretion onto galaxies is SF and it is somewhat paradoxical that, despite our ignorance of the former, the latter is studied in great detail over the lifetime of the Universe.
It is well established that the star formation density in the Universe was higher in the past and has fallen steadily (by a factor about 10) between $z=1$ and now \cite{hippelein+03, hopkins+beacom06}.
Moreover, it is quite clear that star formation histories may differ markedly between the different types of galaxies \cite{panter+07}.
In particular, most blue-cloud galaxies have formed stars at an almost constant rate.

We can learn how galaxies get their gas by studying accretion onto star-forming (blue) galaxies in the local Universe.
A spiral galaxy like the Milky Way, currently forming stars at a rate $\gsim 2 \moyr$ \citep{diehl+06}, needs roughly a similar amount of gas accretion in order not to exhaust its supply in a few Gyrs.
A popular idea is that cold gas clouds, namely High Velocity Clouds (HVCs) are continuously falling onto the Milky Way \cite{oort70, wakker+vanWoerden97, wakker+08}.
However, it is not clear whether the amount of gas brought in by them is sufficient to sustain SF (see below).
In the last decade or so, it has become possible to study anomalous gas clouds and halos in external galaxies and obtain crucial insights into the process of gas accretion \cite{oosterloo+07, sancisi+08}.
Here, I discuss the observations of gas around star-forming disk galaxies and the processes of gas circulation between galaxy disks and halos and viceversa.
I also outline the techniques used to estimate gas accretion rate and show that there is a discrepancy with the star formation rate (SFR).
Finally, I propose 5 possible explanations for this discrepancy.

\section{Halo gas in spiral galaxies}

The presence of gas in the halo region of a spiral galaxy must be evidence of a flow toward or away from the disk.
Halo gas, also called {\it extra-planar gas}, can be produced by several mechanisms.
The most obvious is the so-called galactic fountain or wind\footnote{Note that the terms galactic wind and fountain are both used for ejections of material from the disk but in the former the material is expected to leave the galaxy and be deposited in the intergalactic medium, instead in the latter it will fall back to the disk.} that naturally produces a multiphase thick layer around the star-forming regions \cite{shapiro+field76, bregman80}.
This is the main ``internal'' cause for the extra-planar gas. 
Alternatives have been proposed such as turbulence \cite{struck+smith99} or hydrostatic configurations \cite{barnabe+06} (see also \citet{marinacci+09}).
A completely different category is that of the ``external'' causes.
This includes: i) interactions with other galaxies \cite{yun94}, merging of satellites \cite{sancisi+08}, infall of gas clouds or smooth inflow from the intergalactic medium \cite{ff+07, keres+09}, ram pressure stripping \cite{kenney04}.
In the following, I outline the observational evidence for extra-planar gas in different phases.

\subsection{Hot gas}

Elliptical galaxies are surrounded by halos (also called coronae or atmospheres) of gas at the virial temperature, detected in the X-rays.
Such halos are observed both in massive ellipticals and in galaxies with lower masses.
NGC\,4472 is an example of a large elliptical with total mass $M_{\rm tot} > 10^{12} \mo$ and a hot halo of $M_{\rm gas} > 10^{10} \mo$ and $T_{\rm gas} \sim 2 \times 10^{7} \K$ \citep{irwin+sarazin96}.
On the other hand, NGC\,4697 is a much smaller elliptical, which also has a hot corona of $M_{\rm gas} \sim 1.8 \times 10^{8} \mo$ and $T_{\rm gas} \sim 4 \times 10^{6} \K$ \citep{sarazin+01}.

What about spiral galaxies? A few years ago, the massive spiral galaxy NGC\,5746 was reported by \citet{pedersen+06} to be surrounded by a round halo of hot gas.
Unfortunately, this work has been revisited and it was found that calibration problems had caused a spurious detection and there is no significant diffuse X-ray emission above the plane \citep{rasmussen+09}.
These galactic coronae are in fact not observed in any spiral galaxy to date (see also Mulchaey, this conference) and only indirect estimates exist \citep{fukugita+peebles06}.
The presence of highly ionised atoms in the Milky Way halo is also considered an indirect evidence for this medium \citep{sembach+03}.

However, several spiral galaxies with high SF activity do have extra-planar X-ray emission.
A sample of these galaxies has been studied by \citet{strickland+04} who found typical X-ray extra-planar luminosities of about few times $10^{39}$ erg $s^{-1}$ and temperatures of few times $10^6 \K$.
They also found that the detected emission is proportional to the stellar feedback energy from these galaxies.
From the spatial location of this X-ray emitting gas, the association with the star forming regions is also very clear.
This extra-planar hot gas is therefore a galactic wind or fountain component and \citet{strickland+04} conclude that there is no contribution from IGM material to these halos.

\subsection{Neutral gas}

\begin{figure}[h]
  \includegraphics[width=\textwidth]{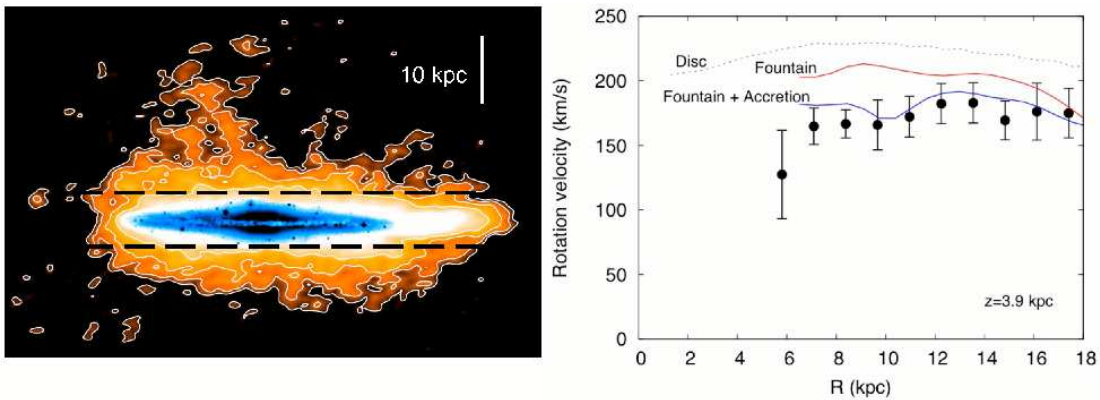}
  \caption{{\it Left:} Total HI map for NGC\,891 (contours) overlaid on an optical image. {\it Right:} rotation curve (points) above the plane of NGC\,891 at an height $z\sim4$ kpc (see dashed lines in the left panel). The prediction of a pure fountain model and of a fountain-driven accretion model are also shown \citep{fb08}.}
\label{f891}
\end{figure}

Neutral gas in nearby disk galaxies is difficult to observe because of its low column density and only recently observations have achieved the required sensitivity \cite{swaters+97, ff+01}.
Extra-planar neutral gas is detected in edge-on galaxies at distances of up to $10-20$ kpc from the plane and, in the case of NGC\,891, it amounts to about 25\% of all the gas in this galaxy (see Fig.\ \ref{f891}).
Table \ref{tExtraplanar} summarizes the best studied galaxies so far.

\begin{figure}[ht]
  \includegraphics[width=\textwidth]{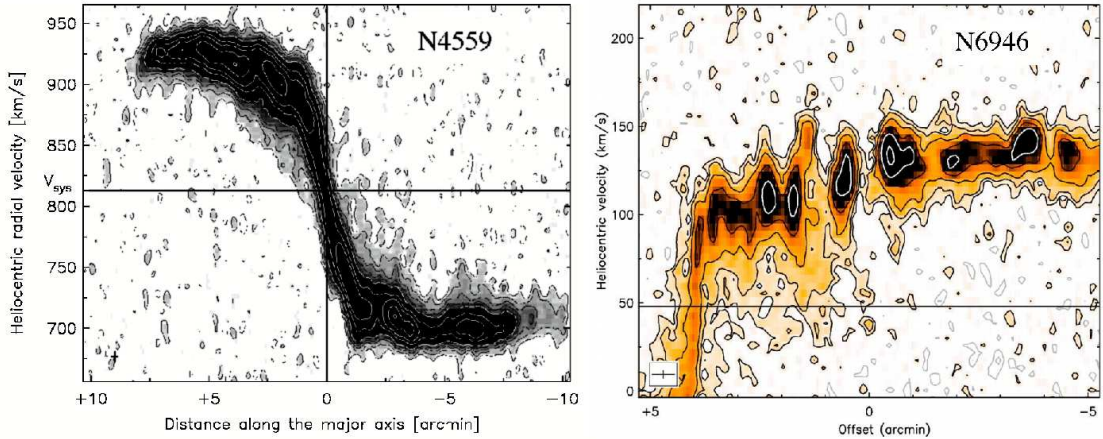}
  \caption{Detection of extra-planar gas in non--edge-on galaxies. The two panels show position-velocity plots along the major axis for the two galaxies NGC\,4559 and NGC\,6946. Most of the emission is detected at velocities close to rotation (roughly the peak emission), however both plots show an ubiquitous one-sided faint tail of emission. This tail is produced by the extra-planar gas, which is rotating more slowly than the disk gas. From \citet{barbieri+05} and \citet{boomsma+08}.}
\label{fBeards}
\end{figure}

The main kinematical feature of the extra-planar gas is its decreasing rotation velocity with increasing height from the plane (vertical rotational gradient). 
Such a velocity gradient has been measured only for a few galaxies (see Table \ref{tExtraplanar}).
In NGC\,891 it has been measured both in \hi\ and in ionised gas giving a consistent value of about $15 \kmskpc$
\citep{oosterloo+07, heald+06_n891, kamphuis+07}.
The existence of such a gradient allows the separation of extra-planar gas from disk gas also in non--edge-on galaxies, Fig.\ \ref{fBeards} shows two examples.
The halo gas shows up at lower rotation velocities with respect to the disk gas and it can be separated assuming a Gaussian velocity profile for this latter.

Given the low temperature and the low kinetic energy of the \hi\ layer, a typical gas cloud is expected to fall down to the disk in less than a dynamical time, typically $0.5-1 \times 10^8$ yr \citep{fb06}.
Thus a mechanism that continuously replenishes the gas in the halo is required.
The most obvious possibility is a galactic fountain and from the amount of gas observed in the halo one can estimate the flow of gas circulating from the disk to the halo and back \citep{collins+02, fb06}.
The chemical evolution of a galactic disk and the spreading of SF throughout it may also be strongly influenced by the disk-halo circulation.

\begin{table}[ht]
\caption{Physical properties of extra-planar neutral gas in spiral galaxies.}
\begin{tabular}{lccccccccc}
\hline
  \tablehead{1}{l}{b}{Galaxy\\}
  & \tablehead{1}{c}{b}{Type\\}
  & \tablehead{1}{c}{b}{i\\ $\de$}
  & \tablehead{1}{c}{b}{v$_{\rm flat}$\\ ${\rm km/s}$}
  & \tablehead{1}{c}{b}{M$_{\rm HI_{\rm halo}}$\\ $\mo$}
  & \tablehead{1}{c}{b}{M$_{\rm HI_{\rm tot}}$\\ $\mo$}
  & \tablehead{1}{c}{b}{SFR\\ $\moyr$}
  & \tablehead{1}{c}{b}{Accr.\\ $\moyr$}
  & \tablehead{1}{c}{b}{Gradient\tablenote{Negative gradient in rotation velocity with height (from the flat part of the rotation curve);}\\ $\rm km/s/kpc$}
  & \tablehead{1}{c}{b}{Ref.\\}\\
\hline
Milky Way &  Sb  & -       & 220      & $\sim4$    & 4   & $1-3$& $\approx0.2$\tablenote{from complex C and other clouds with known distances without correction for the ionised fraction;}& 22\tablenote{extrapolated from the inner 100 pc \citep{levine+08};}& \cite{kalberla+dedes08, levine+08} \\
M\,31     &  Sb  & 77      & 226      & $>0.3$     & 3   & 0.35     &  -  & -          & \cite{thilker+04, walterbos+braun94}\\
M\,33     &  Scd & 55      & 110      & $>0.1$     & 1   & 0.5      &0.05\tablenote{from the HI mass in \citet{grossi+08} without their correction for the ionised fraction;}& -       & \cite{reakes+newton78, grossi+08}\\
M\,83     &  Sc  & 24      & 200      & 0.8        & 6.1 &1.1-2.4 &  -  & -          & \cite{miller+09}\\
NGC\,253  &  Sc  & 75      & 185& 0.8        & 2.5 & $>10$    &  -  & -          & \cite{boomsma+05}\\
NGC\,891  &  Sb  & 90      & 230      & 12         & 4.1 &   3.8    & 0.2 & $15$      & \cite{oosterloo+07} \\
NGC\,2403 &  Scd & 63      & 130      &  3         & 3.2 &   1.3    & 0.1 &$\sim12$\tablenote{not measured directly: derived from the average lag divided by an estimate of the halo thickness;} & \cite{ff+02} \\
NGC\,2613 &  Sb  &  80     & 300& 4.4\tablenote{from the sum of the various extra-planar clouds;}    & 8.7 &   5.1    &  -  & -          & \cite{chaves+irwin01, irwin03} \\
NGC\,2997 &  Sc  & 32      & 226      & 1.4        & 8.0 &   5      & 1.2 & 18-31$^{\rm e}$  & \cite{hess+09} \\
NGC\,3044 &  Sc  & 84      & 150      & 4          & 3   & 2.6\tablenote{calculated from the FIR luminosity using the formula in \citet{kewley02};}  &  -  & -          & \cite{lee+irwin97}\\
NGC\,4559 &  Scd & 67      & 120      & 5.9        & 6.7 & $0.6^{\rm g}$  & $<0.05$  &$\sim10^{\rm e}$ & \cite{barbieri+05}\\
NGC\,5746 &  Sb  & 86      & 310      & $\sim1$    & 9.4 & 1.2      &0.2\tablenote{from the counter-rotating cloud using an infall time-scale of $1\times10^8$ yr;}&  -       & \cite{rand+benjamin08} \\
NGC\,5775 &  Sb  & 86      & 200      &  -         & 9.1 & $7.7^{\rm g}$  &  -  & $8$\tablenote{estimated using optical lines \citep{heald+06_n5775};}     & \cite{irwin94, heald+06_n5775}  \\
NGC\,6503 &  Scd & 75      & 113      & 2-4\tablenote{from the models of \citet{greisen+09}, average between central and outer parts;}    & 1.9 & 0.18     &  -  &$\sim13^{\rm e}$ & \cite{greisen+09} \\
NGC\,6946 &  Scd & 38      & 175      & $\gsim$2.9 & 6.7 &   2.2    &  -  & -          & \cite{boomsma+08}\\
UGC\,7321 &  Sd  & 88      & 110      & $\gsim 0.1$& 1.1 & $\sim0.01$\tablenote{SFR of only massive stars $M> 5 \mo$.}& -&$\lsim25 $& \cite{matthews+03} \\
\hline
\end{tabular}
\label{tExtraplanar}
\end{table}

\subsection{Ionised gas}

The extra-planar gas is also observed in optical emission lines.
These are often called diffused ionised gas (DIG) layers and they extend up to several kpc from the plane \citep{hoopes+99, collins+00}.
The most likely source of ionization is the radiation flux from the stars in the disk \citep{reynolds90}, however it is not clear whether this source is sufficient \citep{rand+08}.
\citet{rossa+dettmar03} studied a sample of 74 edge-on galaxies in optical emission lines and found that about 40\% have significant extra-planar emission.
The kinematics of the extra-planar ionised gas is similar to that of the neutral gas.
The gradient in rotational velocity has been estimated for NGC\,891 (see above) and also for NGC\,5775 and NGC\,4302 in the ionised gas \citep{heald+06_n5775, heald+07}.
In NGC\,2403, \citet{ff+04} found tails at low rotational velocities in the ionised gas profiles very similar to those found in the neutral gas.

\section{Direct detection of gas accretion}
\label{accretion}

\begin{figure}[ht]
  \includegraphics[width=0.45\textwidth]{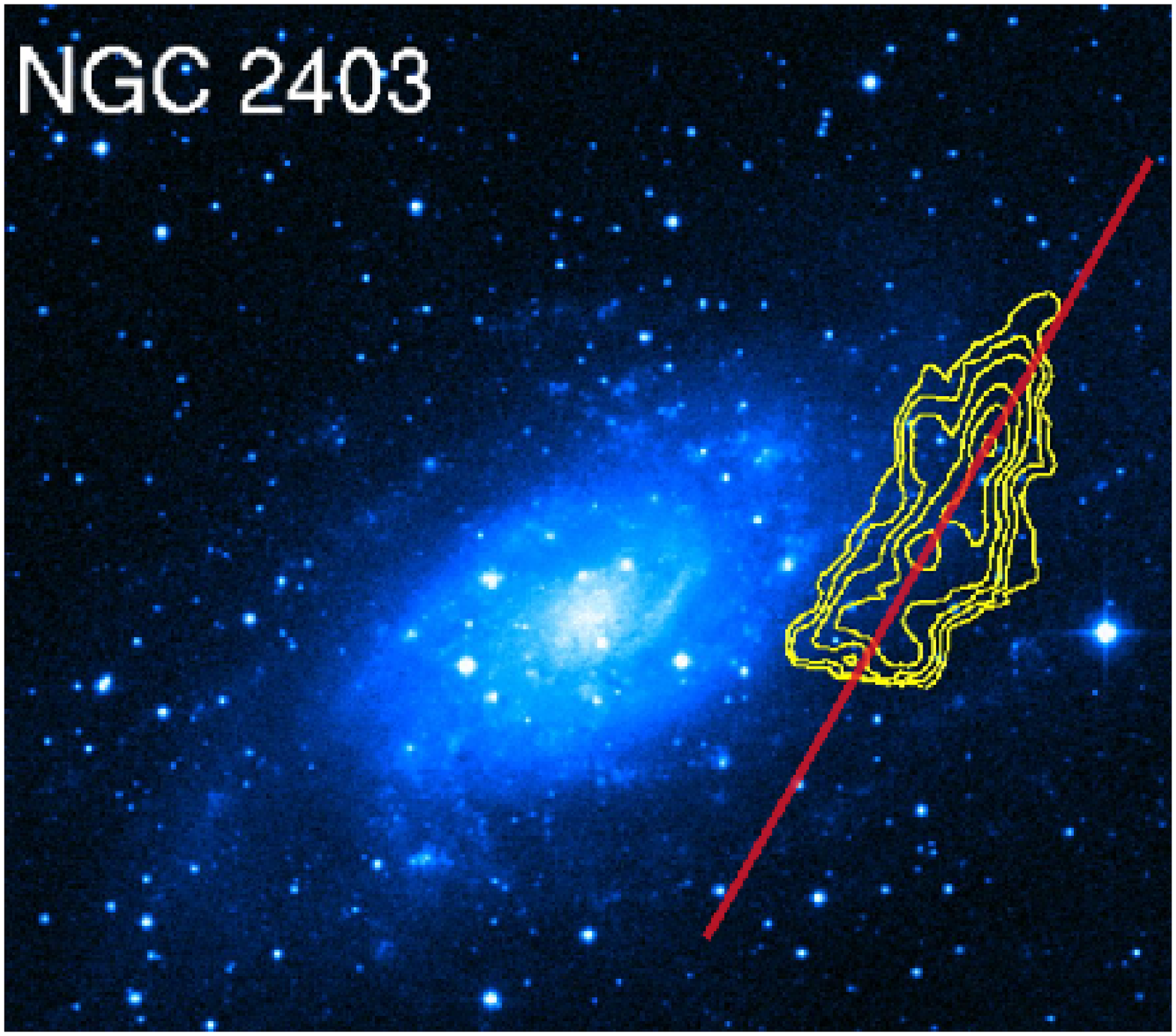}
  \includegraphics[width=0.51\textwidth]{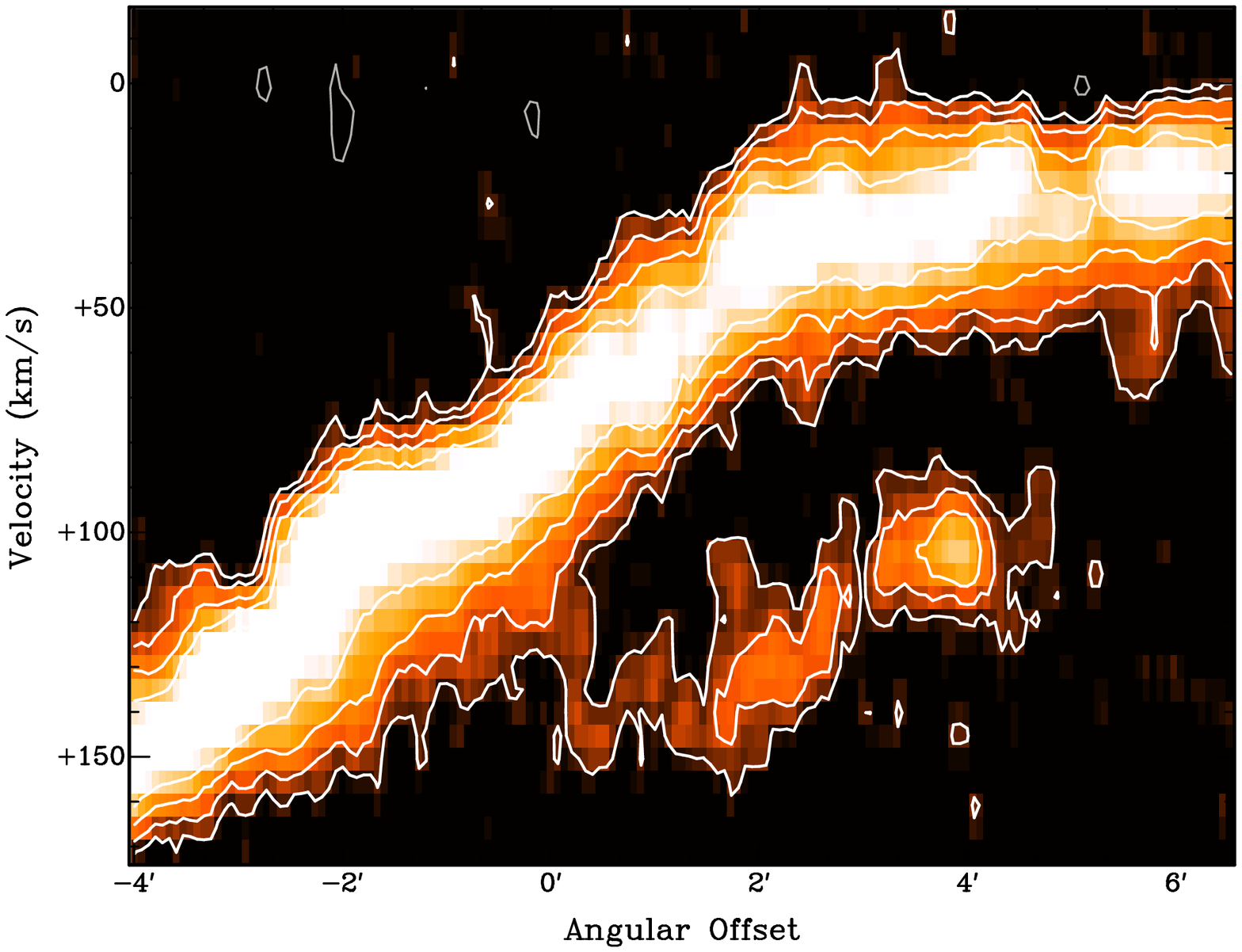}
  \caption{A massive HI complex (filament) in NGC\,2403. 
{\it Left:} optical image overlaid with the HI contours of the $1 \times 10^7 \mo$ filament. 
{\it Right:} position-velocity plot along the line in the left panel. The filament is 8 kpc long and it has kinematics clearly separated from the normal disk gas. From its mass one can estimate an accretion rate of $\sim 0.1 \moyr$.}
\label{f2403}
\end{figure}

The strategy to estimate halo-gas accretion is to look for gas components (usually at very anomalous velocities)
which are incompatible with an internal (galactic fountain) origin.
The large majority of the extra-planar gas studied so far has actually very regular kinematics that follows closely the kinematics of the disk (see for instance the p-v diagrams in Fig.\ \ref{fBeards}).
However, there are features that do not obey this rule.
NGC\,891 has a long filament ($M_{\rm HI} \sim 1.6 \times 10^7 \mo$) extending up to about 20 kpc from the plane of the disk (Fig.\ \ref{f891}).
NGC\,2403 also has a filament with a similar \hi\ mass located in projection outside the bright optical disk (Fig.\ \ref{f2403}).
They are both very similar to Complex C in our Galaxy \citep{wakker+07}.
The energy needed to form these complexes assuming that they are originated by a galactic fountain is of the order $\sim 1 \times 10^{55}\,$erg corresponding to the {\it simultaneous} explosion of about $10^5$ supernovae, a very unlikely event.
Other features that deserve attention in the search for gas accretion are clouds at very anomalous velocities that appear to counter-rotate \citep{ff09}.
These clouds cannot be produced in any kind of galactic fountain and they are likely to be direct evidence of gas accretion.

Using the masses of the accretion structures quoted above, we can estimate the rate of gas accretion by assuming typical infall times of $1\times 10^8\,$yr.
The resulting rates are shown in Table \ref{tExtraplanar} (column 8), they are typically of 
the order $0.1 \moyr$ and generally 1 order of magnitude lower than the SFRs.
For NGC\,4559 there is no obvious evidence for accretion of the kind described above as all the gas at anomalous velocity seems connected to the disk gas.
Therefore I estimate a fiducial upper limit of order $0.05 \moyr$.
The directly observed accretion rates shown in Table \ref{tExtraplanar} include only \hi, and they should be corrected for helium and possibly ionised gas fractions (see below).

\section{The {\it hidden} gas accretion}

Comparing the columns 7 and 8 in Table \ref{tExtraplanar},
there seems to be a discrepancy between the directly measured gas accretion and what would be required to mantain the SFR observed in these galaxies.
The uncertainties in the determination of the accretion rates are large as we discuss below, however the gap that needs to be filled is of one order of magnitude.
Thus, it appears that most of the accretion is ``hidden'' from our current investigations.
I consider 5 possible explanations.\\

{\bf 1. There are other $\hi$ clouds that have escaped detection}

Several observations of the fields around galaxies and of groups similar to the Local Group have been undertaken \citep{pisano+04, pisano+07, chynoweth+09} with detection limits of $\lsim 10^6 \mo$.
No floating clouds have been detected.
Large blind surveys, such as HIPASS \citep{barnes+01} and the ongoing large survey ALFALFA also found no evidence for a significant population of isolated floating \hi\ clouds (so-called dark galaxies) in the IGM. 
The first results from ALFALFA show that only 3\% of \hi\ sources are not detected in the optical \citep{giovanelli+07} although a populations of low-mass clouds may be present in the Local Group \citep{giovanelli+10}.
These results also agree with what is known from the very deep \hi\ observations of some nearby galaxies, like those of M\,31 \citep{thilker+04} and NGC\,891 \citep{oosterloo+07}.
These data show \hi\ clouds with masses of $\gsim 1 \times 10^6 \mo$ located very close (a few tens of kpc) to the host galaxy. 
Therefore, there is no evidence for an IGM population of neutral gas clouds capable of fuelling the SF.
Possibly, one should think in terms of a {\it drizzle} \citep{fb08} or smooth accretion, virtually undetectable in emission (see Katz, this conference).\\

{\bf 2. Most of the accreting gas is ionised}

As mentioned, a fraction of the extra-planar gas is ionised but it is not clear how important this fraction is, given the uncertainties in estimating the gas mass from optical emission lines.
Interestingly, the fraction of extra-planar neutral gas in starburst galaxies, see NGC\,253 for instance \citep{boomsma+05}, seems lower than in other galaxies (see Table \ref{tExtraplanar}) showing that in these cases most gas probably is ionised.
In general, in a galactic fountain scheme, it is reasonable to assume that the fraction of ionized gas is at most equal to that of neutral gas \citep{fb06}.

Clearly, also a fraction of the accreting gas will be ionised.
In this case, a second important cause for ionization is the interaction of the infalling cloud with the corona.
There have been attempts to estimate this fraction for HVCs \cite{tufte+98} that led to the conclusion that the mass of ionised gas is roughly comparable to that of neutral gas \citep{wakker+07}.
This would produce a correction of a factor 2 in the gas accretion values of Table \ref{tExtraplanar}.
Larger corrections have been applied in some cases, the most remarkable being the study of M\,33 by \citet{grossi+08} where they applied a factor of up to 50, but such assumptions appear hard to justify.

Even if the discrepancy could be filled with ionised gas the question remains of how gas coming from the intergalactic medium and thus originally in a hot phase can cool and become star-forming gas.
If spiral galaxies are embedded in massive coronae a mechanism is needed for this medium to cool and fall into the disk.
This process is not understood because, if on the one hand, the cooling time can in principle be short enough, on the other hand perturbations are likely to be suppressed very easily.
\citet{binney+09} show that a combination of buoyancy and thermal conduction causes thermal stability for a large variety of coronal parameters (see Nipoti, this conference).
Simulations of cooling coronae also predict low accretion rates of order $0.2-0.3 \moyr$ even in the most favorable cases \citep{kaufmann+09, peek+08}.
Therefore, a different mechanism is required for transferring gas from the IGM to the galactic disks (see point 5).\\

{\bf 3. Minor mergers and interactions}

\begin{figure}[ht]
  \includegraphics[width=0.49\textwidth]{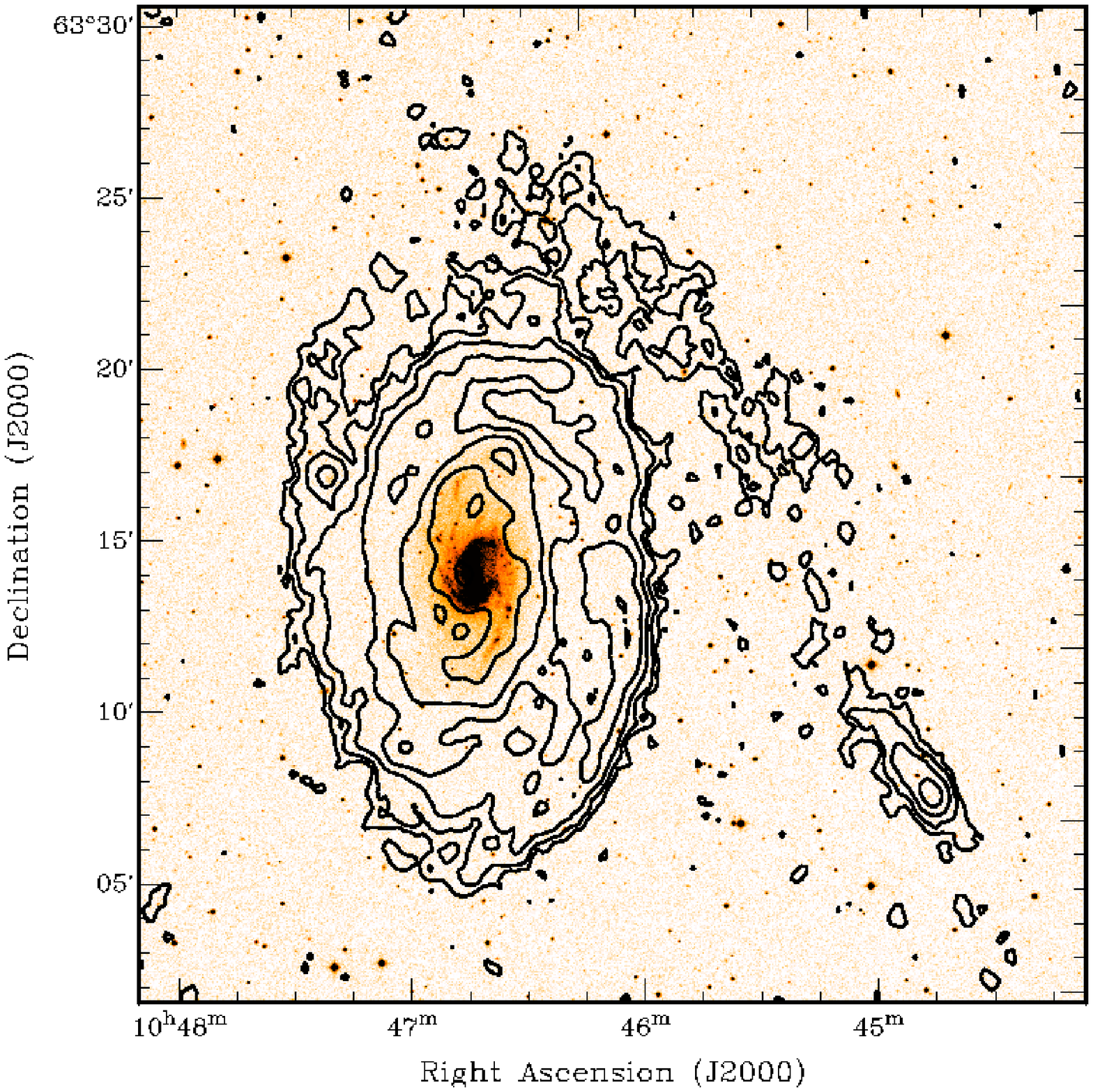}
  \includegraphics[width=0.54\textwidth]{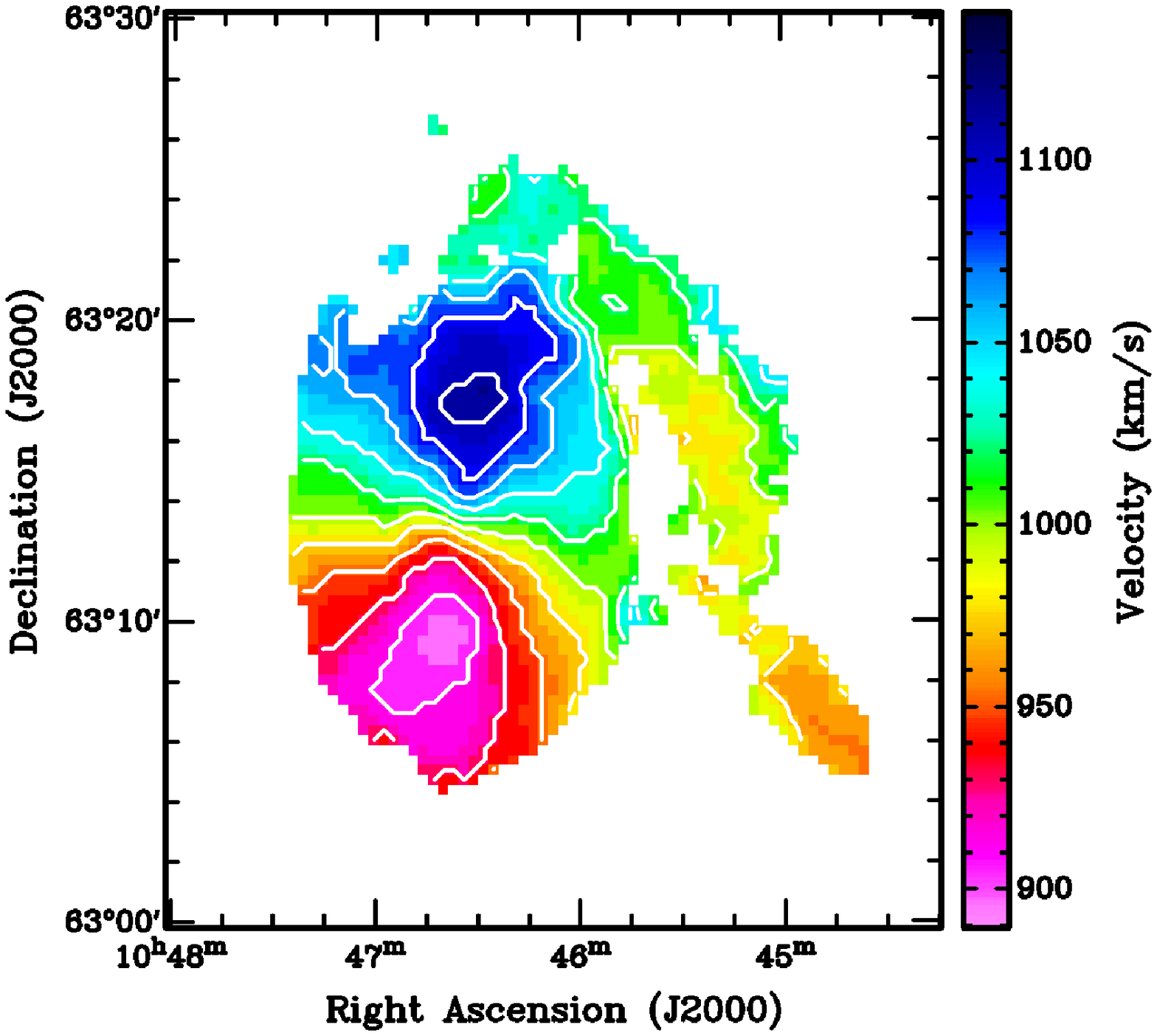}
  \caption{NGC\,3359 in the process of swallowing a gas rich companion. {\it Left:} total HI map (contours) overlaid on an optical image, {\it right:} HI velocity field.}
\label{fMerging}
\end{figure}

Contribution to the gas accretion can also come from the merging of gas-rich satellites, an example of such process is shown in Fig.\ \ref{fMerging}.
A rough estimate of the amount of gas brought in by satellites in Milky-Way types of galaxies has been obtained by \citet{sancisi+08} using the WHISP catalogue.
They found a value of about $0.1 - 0.2 \moyr$, compatible with the above estimates. 
Such a value is very uncertain and it could be severely underestimated.
A way to test this is to check whether there would be enough gas-rich dwarf galaxies to feed larger galaxies at the required rate. 
The \hi\ mass function shows that most of the neutral gas in the local Universe is in large galaxies, with masses above $\sim 1 \times 10^9 \mo$ \citep{zwaan+05}. 
If we assume that large galaxies accrete gas only via minor mergers and we impose an accretion rate of $1 \moyr$, then all small galaxies would be accreted by the large galaxies (and disappear) on a timescale of the order of a Gyr. 
Thus, constant gas accretion rates as high as the current SFR cannot be achieved via mergers with satellite galaxies simply because there are not enough dwarf galaxies to be accreted \citep{sancisi+08}. 
This is a major problem for this class of solutions, the only way out being that the faint end of the \hi\ mass function was much steeper in the past.
This possibility will be tested in the next few years with the SKA pathfinders.\\

{\bf 4. There is no need for accretion: SF is dying out}

A common criticism to the discrepancy between directly measured gas accretion and SF rates in the local Universe is that SF is coming to a halt.
In this view, galaxies in the local blue cloud are just exhausting the gas that they have been given from the beginning without a significant contribution of accretion from the IGM.
A consequence of this reasoning is, of course, that the amount of gas in galaxies must have been much larger at high redshifts.

At present, observations of the so-called Damped Lyman Alpha (DLA) systems, long recognized to be the high-z analogues of the local population of (gas-rich) spiral galaxies are the only way to study neutral gas in galaxies at high z.
A comparison between DLA studies and local galaxy surveys such as HIPASS indicates that the amount of neutral gas in galaxies has not changed significantly throughout the Hubble time, at most it has declined by a factor less than 2 between $z=4$ and the present \citep{zwaan+05, prochaska+05}.
This is consistent with the fact that the SFR in the Milky Way has not changed substantially over the same timeframe \cite{twarog80, cignoni+06}. 
The behaviour of the gas mass in galaxies is remarkably different to that of the stellar mass: between $z=4$ and now the stellar mass has built up by a factor more than 10 in a galaxy like the Milky Way.
This is very likely the effect of gas accretion over time \citep{hopkins+08}.
A second possibility is that the vast majority of gas in the past was in molecular form but this leads to unrealistic gas densities and molecular gas consumption rates in disagreement with the observations \citep{bauermeister+09}.\\

{\bf 5. Fountain-driven gas accretion}

The steepness of the vertical rotational gradient is not reproduced by galactic fountain models \citep{fb06} as they tend to predict shallower values (a factor half or less).
Fig.\ \ref{f891} (right panel) highlights this problem for NGC\,891, the points are rotation velocities derived at the height of $z=3.9\,{\rm kpc}$ from the plane.
Clearly the fountain clouds in the model rotate too fast (have a larger angular momentum) than the observed extra-planar gas.

\begin{figure}[ht]
  \includegraphics[width=\textwidth]{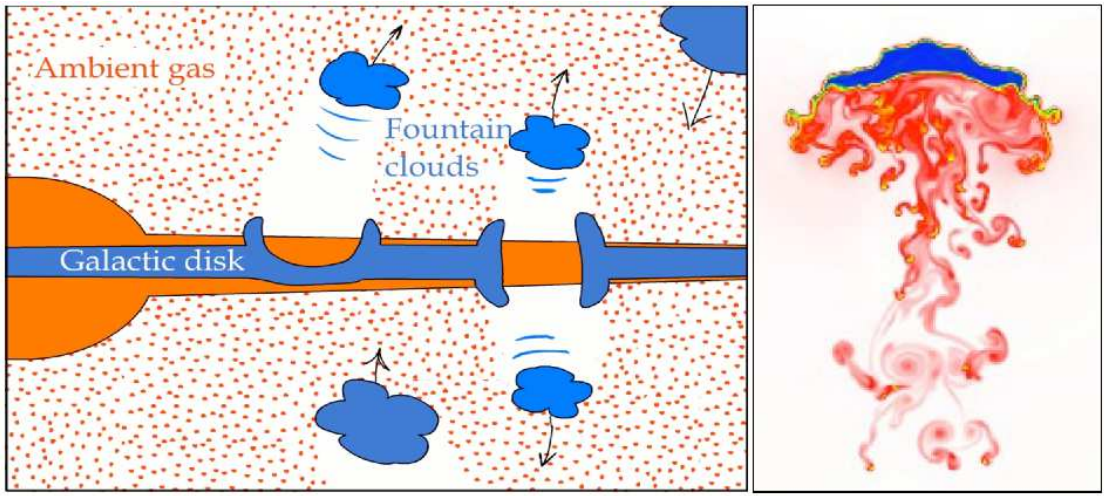}
  \caption{{\it Left:} schematic view of the fountain-driven accretion process. Gas clouds leaving the disk sweep up ambient gas travelling through the galactic corona \citep{ff09}.
{\it Right:} hydrodynamical simulation of a cold cloud moving through a hot medium. The condensation of coronal gas occurs in the turbulent wake of the cloud \citep{marinacci+10}.}
\label{fFB08}
\end{figure}

\citet{fb08} consider the possibility that fountain clouds sweep up ambient gas as they travel through the halo.
In this scheme ambient gas condenses onto the fountain clouds, which grow along their paths through the halo and eventually fall down into the disk (see Fig.\ \ref{fFB08}, left).
If the ambient gas has relatively low angular momentum about the z-axis then this process produces a reduction in the rotational velocity of the fountain gas.
The only free parameter of the model is the accretion rate, which is tuned to reproduce the rotation curves of the extra-planar gas.
Remarkably, the required gas accretion rate turns out to be very similar to the SFR.
For NGC\,891 we found a best-fit accretion rate of about $3 \moyr$  (see curve in Fig.\ \ref{f891}). 
This model also predicts that most of the extra-planar gas is produced by the galactic fountain and only a small fraction (about 10\%) is extragalactic.

The physical mechanism for a fountain cloud to accrete ambient material is not yet established.
In recent hydrodynamical simulations (Fig.\ \ref{fFB08}, right), \citet{marinacci+10} explored the possibility that condensation of coronal gas occurs in the turbulent wakes of fountain clouds (see also Marinacci et al., this conference).
It was found that, for realistic assumptions for the fountain clouds and the galactic corona properties, condensation is likely to be efficient enough to produce an accretion rate of the required amount.
Thus, a process of this kind is a viable mechanism for star-forming galaxies to acquire fresh (low metallicity) gas from the IGM; a somewhat similar mechanism has been proposed by \citet{oppenheimer+10}.
If the hot corona cools efficiently in the cloud wakes, one would expect a significant amount of gas in the temperature range between $10^4$ and $10^6$ K around HVCs and IVCs.
The highly ionised material mapped by FUSE \citep{sembach+03} and the newly detected widespread population of Si III and Si IV absorbers \citep{shull+09} are potentially evidence of this kind of ongoing phenomenon.
Taken in connection with the declining efficiency of cold mode accretion \citep{dekel+birnboim06},
this fountain-driven accretion could be the dominant accretion mode for star-forming galaxies at redshifts $z\lsim1$.




\bibliographystyle{aipproc}   

\bibliography{filippo}

\IfFileExists{\jobname.bbl}{}
 {\typeout{}
  \typeout{******************************************}
  \typeout{** Please run "bibtex \jobname" to optain}
  \typeout{** the bibliography and then re-run LaTeX}
  \typeout{** twice to fix the references!}
  \typeout{******************************************}
  \typeout{}
 }

\end{document}